\documentclass{article}

\usepackage{times}
\usepackage{graphicx} 
\usepackage{subfigure}

\usepackage{natbib}

\usepackage{algorithm}
\usepackage{algorithmic}

\usepackage{pygtex}

\usepackage{hyperref}

\usepackage[accepted]{data4good2016}

\icmltitlerunning{Harnessing the Power of the Crowd to Increase Capacity for Data Science in the Social Sector}

\begin{document}

\twocolumn[
\icmltitle{Harnessing the Power of the Crowd to \\
  Increase Capacity for Data Science in the Social Sector}

\icmlauthor{Peter Bull}{peter@drivendata.org}
\icmladdress{DrivenData, Cambridge, MA USA}
\icmlauthor{Isaac Slavitt}{isaac@drivendata.org}
\icmladdress{DrivenData, Cambridge, MA USA}
\icmlauthor{Greg Lipstein}{greg@drivendata.org}
\icmladdress{DrivenData, Cambridge, MA USA}

\icmlkeywords{Machine learning, statistical modeling, crowdsourcing, nonprofits, governments, education, public health, government innovation}

\vskip 0.3in
]

\begin{abstract}
We present three case studies of organizations using a data science competition to answer a pressing question. The first is in education where a nonprofit that creates smart school budgets wanted to automatically tag budget line items. The second is in public health, where a low-cost, nonprofit womens' health care provider wanted to understand the effect of demographic and behavioral questions on predicting which services a woman would need. The third and final example is in government innovation: using online restaurant reviews from Yelp, competitors built models to forecast which restaurants were most likely to have hygiene violations when visited by health inspectors.\footnote{These competitions were run on the DrivenData competition platform (www.drivendata.org); DrivenData employs the authors of this paper.} Finally, we reflect on the unique benefits of the open, public competition model.
\end{abstract}

\begin{figure}[ht]
\vskip 0.2in
\begin{center}
\centerline{\includegraphics[width=\columnwidth]{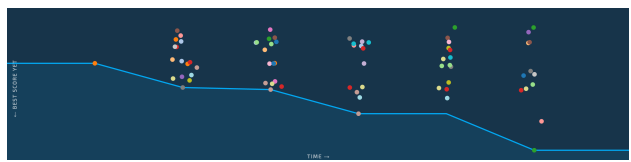}}
\caption{The best predictions (lower is better) over time in a data science competition. Different color dots represent different submitters.}
\label{icml-historical}
\end{center}
\vskip -0.2in
\end{figure}

\section{Introduction}
\label{submission}
If your goal is to change the future, it helps to have good predictions about what that future looks like.

And there are many groups interested in changing the future. Amazon wants to increase the number products you order, so they predict which ones you might want to buy next and recommend them.\cite{amazon} Twitter wants to boost your use of their platform, so they predict which tweets you will ignore and which you will engage with.\cite{lin2012large} Facebook and Google want to increase the number of ads you click on their sites, so they predict your personal click-through behavior.\cite{google} These companies have become extraordinarily skilled at making these predictions.

But there are many other reasons to want to change the future. Educators want to increase the number of students graduating high school. Health workers want to improve the overall health of a population at a sustainable cost. Microlenders want to give more individuals in the developing world a chance to pursue their dreams without incurring default. Conservationists want to curb our energy usage without hampering productivity. Governments want to prevent fires from destroying lives and property.

We have the computational power and methods to tackle many of these challenges. However, the data scientists that can leverage these resources are hard to find and expensive to hire. McKinsey has estimated that in 2018 there will be 190,000 analytics positions that go unfilled in the United States.\cite{mgi} If there is that kind of shortfall in the commercial sector, we expect the social sector will lag even further behind. In 2014, the median salary for a data scientist was \$98,000, which puts qualified data scientists out of reach as full time hires at many nonprofits.\cite{dsss} In the last ten years, open innovation has been recognized as a cost-effective way to generate creative, high-quality solutions to problems.\cite{boudreau2013using} Data science competitions, where practitioners, researchers, and students publicly compete to make the best predictions, offer a way to harness the power of open innovation for data analysis.

Competitions have great visibility among both data science practitioners and other nonprofits working in similar issue areas. Participating in a competition spurs data scientists to think more about applicable problems in the social sector, which is especially important at a time where creative applications of new data techniques can greatly improve how organizations operate. These competitions also help start the conversation among nonprofits about how machine learning tools can be used on their own data.

\section{Case Study: ``Box-plots for Education''}

\subsection{Context}
Education Resource Strategies (ERS) was founded as a non-profit consulting firm in 2004, with a primary goal of helping public school districts use their limited resources more strategically. However, this goal is often easier said than done. Unlike companies, which benefit from comparing themselves to their peers, districts frequently have no reliable ways to compare their spending to other districts. Even if expenditures are public, the comparisons are not apples-to-apples because of a lack of standardized reporting structures. As a result, district decision makers are often left in the dark.

ERS attempts to solve this problem by working with districts' budgets to assign every line item to certain categories in a comprehensive financial spending framework. If this process is completed correctly, ERS can offer cross-district insight into a partner district's finances. For example, ERS might observe that a particular district spends more on facilities and maintenance than peer districts--while this is not inherently good or bad, it helps the district to have knowledge of how its decisions and spending compare to those of its peers.

In order to compare budget or expenditure data across districts, ERS assigns every line item to certain categories in a comprehensive financial spending framework. For instance, some labels describe what the spending ``is"--compensation, benefits, equipment, property rental, and so on. Other categories describe what the spending ``does," which groups of students benefit, and where the funds come from.

However, categorizing each of these budget line-items is extremely labor-intensive, often taking several weeks for an employee to hand-tag each row in Excel. This challenge put a limit on the quality of comparisons, because ERS could only process so many budgets per year.

\subsection{Challenge}
In some senses, this is a classic machine learning problem. ERS had hand-labeled over 450,000 budget line items with the relevant categories. The objective was to train the machine to do the process that they had been doing by hand. This kind of problem is known as supervised machine learning.

In this competition, competitors were asked to identify the highest probability labels for each of nine different categories. To do this, they had to first create features from the text of the district budgets. We include some examples of text from school budgets to demonstrate that this is not a standard natural language processing task--in fact, it is very much ``unnatural'' language processing given the number and variety of abbreviations and punctuation.

\begin{table}
\small
\centering
\caption{Example text from school district budgets}
\vspace{2 mm}
\texttt{
PETRO-VEND FUEL AND FLUIDS\\
Regional Playoff Hosts\\
Capital Assets - Locally Defined Groupings\\
FURNITURE AND FIXTURES\\
ITEMGH EXTENDED DAY \\
Water and Sewage * \\
UPPER EARLY INTERVENTION PROGRAM 4-5 \\
Food Services - Other Costs \\
Supp.- Materials \\
}
\vspace{2 mm}
\end{table}

\subsection{Results}
The winning algorithm came from a competitor who had submitted over 100 times to the competition--a testament to the passion of data scientists working for a cause. The algorithm uses a standard method--logistic regression--but derives its power from its feature engineering: for example, using tri-grams, pairwise interactions, the ``hashing trick'' for dimensionality reduction, and term frequency-inverse document frequency (tf-idf) among other techniques.

ERS estimates that this algorithm will tag files with over 90\% accuracy and will save 75\% of the time usually taken to code financial files. At 400 hours per project, this means 300 hours saved per project, or close to 1,000 hours per employee, who typically does three projects per year. Ultimately, this equates to roughly twenty-five weeks of employee time saved!

\section{Case Study: ``Countable Care''}

\subsection{Context}
Planned Parenthood is the nation's leading provider and advocate of high-quality, affordable healthcare for women, men, and young people, as well as the nation's largest provider of sex education. With approximately 700 health centers across the country, Planned Parenthood organizations serve all patients with care and compassion, with respect and without judgment. Understanding the trends in women's health care is critical to delivering the expert, quality care that is the hallmark of Planned Parenthood.

Planned Parenthood is an innovator in health care delivery, continually looking to find the best ways to expand access to quality, affordable care to everyone who needs it. We want your help to better understand the complex dynamics of health care in order to better serve the needs of those who depend on us.

The goal of this competition is to drive innovation and analysis in the field of population health by predicting which reproductive health care services are accessed by women. The end product of the competition will improve public health with novel predictive analytics as part of our effort to give back to the research and healthcare community.

\subsection{Challenge}

Users were given extremely detailed information from multiple years of the Centers for Disease Control and Prevention (CDC) reproductive health and family choice survey, the National Survey of Family Growth (NSFG). The survey ``gathers information on family life, marriage and divorce, pregnancy, infertility, use of contraception, and men's and women's health'' and is widely used by researchers focusing on contraception, family planning, and women's health in general.\cite{chandra2005fertility}

In addition to a vast amount of in-depth demographic and personal history information, the NSFG tracks captures fairly detailed family and reproductive histories, as well as health decisions. We challenged users to model deeply latent, non-linear, and non-obvious associations between demographics and personal histories in order to predict these health choices for individual respondents.

This is a particularly interesting challenge given the branching nature of the survey. Because respondents were only asked certain questions if they answered affirmatively to a previous question, there was a lot of missing data in the dataset. It was an open challenge to competitors to determine how to treat this missing values and still make effective predictions.

\begin{figure}[ht]
\vskip 0.2in
\begin{center}
\centerline{\includegraphics[width=\columnwidth]{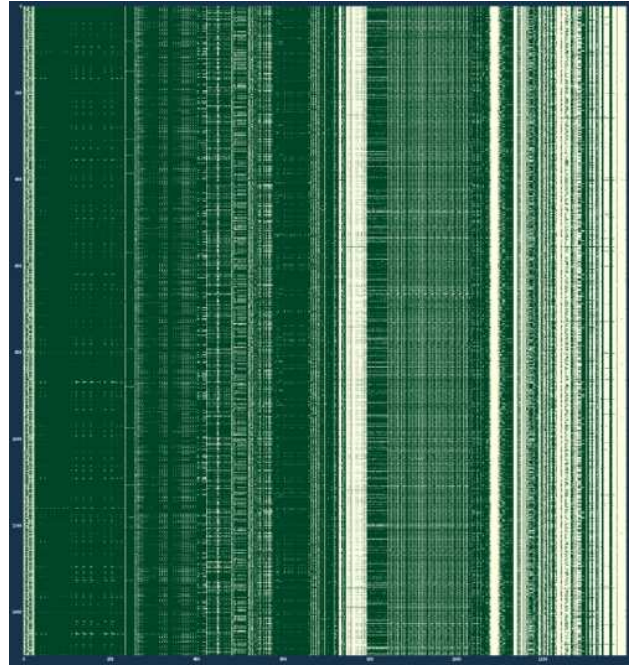}}
\caption{Green pixels are questions that were answered; yellow pixels were gaps in the data where survey questions were unanswered.}
\label{nans}
\end{center}
\vskip -0.2in
\end{figure}

\subsection{Results}

Using robust ensembles of individually sophisticated models, the top performers in this challenge were able to achieve significant predictive lift over both random noise and more naive models. The code and models associated with this challenge were delivered to the Guttmacher Institute for further study.

\section{Case Study: ``Keeping it Fresh''}

\subsection{Context}
The City of Boston regularly inspects every restaurant to monitor and improve food safety and public health. As in most cities, health inspections are generally random, which can increase time spent on spot checks at clean restaurants that have been following the rules closely--and missed opportunities to improve health and hygiene at places with more pressing food safety issues.

Each year, millions of people cycle through and post Yelp reviews about their experiences at these same restaurants. The information in these reviews has the potential to improve the City's inspection efforts, and could transform the way inspections are targeted.

A team of Harvard economists and Yelp--with support from the City of Boston--co-sponsored this competition to explore ways to use Yelp review data to improve the inspections process. The City of Boston, a partner in this challenge, was committed to examining ways to integrate the winning algorithm into its day-to-day inspection operations.

\subsection{Challenge}

The goal for this competition was to use data from social media to narrow the search for health code violations in Boston. Competitors were given access to historical hygiene violation records from the City of Boston and a massive archive of Yelp's consumer reviews along with restaurant metadata. Specifically, users were directed to predict the results of an inspection of every restaurant on every day for the time period in question. The challenge is to figure out the words, phrases, patterns in foot traffic, prices, cuisines, and other clues that harness digital exhaust to make city services more effective.

\begin{table}
\small
\centering
\caption{Example review from over 230k in the Yelp dataset}

\begin{Verbatim}[commandchars=\\\{\},frame=leftline,framesep=1.5ex,framerule=0.8pt]
\PY{p}{\PYZob{}}
  \PY{n+nt}{\PYZdq{}business\PYZus{}id\PYZdq{}}\PY{p}{:} \PY{l+s+s2}{\PYZdq{}CgdK8DiyX9Y4kTKEPi\PYZus{}qgA\PYZdq{}}\PY{p}{,}
  \PY{n+nt}{\PYZdq{}type\PYZdq{}}\PY{p}{:} \PY{l+s+s2}{\PYZdq{}review\PYZdq{}}\PY{p}{,}
  \PY{n+nt}{\PYZdq{}text\PYZdq{}}\PY{p}{:} \PY{l+s+s2}{\PYZdq{}This is the place I like to go}
\PY{l+s+s2}{    for deli sandwiches (and salads/soups)}
\PY{l+s+s2}{    when in the FinancialDistrict. I\PYZsq{}m not}
\PY{l+s+s2}{    sure what makes this place stand out}
\PY{l+s+s2}{    from the million other deli sandwich}
\PY{l+s+s2}{    places inthe area. Maybe it\PYZsq{}s the lack}
\PY{l+s+s2}{    of pretentiousness...\PYZdq{}}\PY{p}{,}
  \PY{n+nt}{\PYZdq{}date\PYZdq{}}\PY{p}{:} \PY{l+s+s2}{\PYZdq{}2005\PYZhy{}12\PYZhy{}11\PYZdq{}}\PY{p}{,}
  \PY{n+nt}{\PYZdq{}stars\PYZdq{}}\PY{p}{:} \PY{l+m+mi}{4}\PY{p}{,}
  \PY{n+nt}{\PYZdq{}review\PYZus{}id\PYZdq{}}\PY{p}{:} \PY{l+s+s2}{\PYZdq{}zQH071b6x9g1ZHbhJnaNKw\PYZdq{}}\PY{p}{,}
  \PY{n+nt}{\PYZdq{}user\PYZus{}id\PYZdq{}}\PY{p}{:} \PY{l+s+s2}{\PYZdq{}NfvN6\PYZhy{}zeU0RsD0Q\PYZus{}Sk\PYZhy{}DSQ\PYZdq{}}\PY{p}{,}
  \PY{n+nt}{\PYZdq{}votes\PYZdq{}}\PY{p}{:} \PY{p}{\PYZob{}}
    \PY{n+nt}{\PYZdq{}cool\PYZdq{}}\PY{p}{:} \PY{l+m+mi}{1}\PY{p}{,}
    \PY{n+nt}{\PYZdq{}useful\PYZdq{}}\PY{p}{:} \PY{l+m+mi}{1}\PY{p}{,}
    \PY{n+nt}{\PYZdq{}funny\PYZdq{}}\PY{p}{:} \PY{l+m+mi}{0}
  \PY{p}{\PYZcb{}}
\PY{p}{\PYZcb{}}
\end{Verbatim}

\end{table}

\begin{figure}[ht]
\vskip 0.2in
\begin{center}
\centerline{\includegraphics[width=\columnwidth]{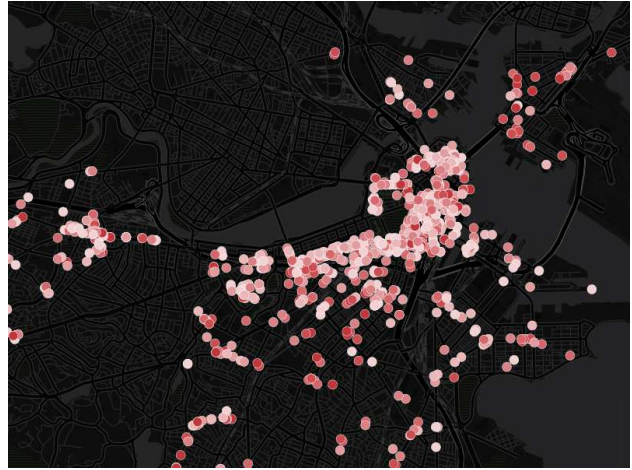}}
\caption{A map of hygiene violations in the City of Boston. Darker circles have more violations historically, while lighter ones have fewer.}
\label{bos}
\end{center}
\vskip -0.2in
\end{figure}

\subsection{Results}

The competition opened Monday, April 27th 2015 and accepted normal submissions for eight weeks. During this period, users could see how well their predictions were relative to the holdout data, and compare their performance with other competitors on the public leaderboard.

In this competition, data scientists were not simply trying to make predictions on normal holdout data alone. Rather, they were challenged actually to predict the future. In addition to a normal public leaderboard, this competition featured an evaluation period after normal submissions closed; prior to submissions being closed, competitors made special submissions not on the holdout dataset but on the upcoming six weeks of actual Boston food inspections.

By July 7th 2015, normal submissions were closed and users had submitted their final evaluation predictions for the evaluation period. Over the next six weeks, as the city of Boston went about their normal inspection routines we pulled the violations from the open data portal and evaluated competitors performance in realtime. At the end of the competition, the results were evaluated by a team of researchers at Harvard University who ``estimate that the City of Boston would be 30\%-50\% more productive using a top-performing algorithm from the tournament . . . [and] are currently testing the winning algorithms’ efficacy in practice, using a field experiment that integrates the winning algorithms into Boston’s process for allocating inspectors.'' \cite{NBERw22124}

\section{Conclusion}

It is an exciting time to be working on data-for-good projects. With a bit of creativity, data scientists can create analogies between problems that are being solved in industry and the challenges facing nonprofits, NGOs, and governments. The tools that are used by corporations to improve their operations and bottom-lines can just as easily be used to help social impact organizations be more effective and more efficient.

Open innovation provides a new way for nonprofits and governments to access talent that is hard to find and expensive. Experts from around the world can contribute to social impact from wherever they are, whenever they have free time. Nonprofits can be almost guaranteed high performing algorithms given the sheer number of models explored during a competition. Both groups can learn from and explore new applications of existing techniques to make the world a better place.

\bibliography{drivendata_icml.bib}
\bibliographystyle{icml2016}

\end{document}